\pdfoutput=1
\documentclass[12pt,a4paper,titlepage]{article}
\usepackage[utf8]{inputenc}
\usepackage[top=50pt,bottom=50pt,left=68pt,right=66pt]{geometry}
\usepackage{amsmath}
\usepackage{booktabs}
\usepackage{amssymb}
\usepackage[title]{appendix}
\usepackage{mathrsfs}
\usepackage{float}
\usepackage{multirow}
\usepackage{graphicx,caption,subcaption}
\usepackage[space]{grffile}

\begin{document}
\renewcommand{\arraystretch}{1.3}

\makeatletter
\def\@hangfrom#1{\setbox\@tempboxa\hbox{{#1}}%
      \hangindent 0pt
      \noindent\box\@tempboxa}
\makeatother


\def\un#1{\relax\ifmmode\@@underline#1\else
        $\@@underline{\hbox{#1}}$\relax\fi}


\let\under=\unt                 
\let\ced=\ce                    
\let\du=\du                     
\let\um=\Hu                     
\let\sll=\lp                    
\let\Sll=\Lp                    
\let\slo=\os                    
\let\Slo=\Os                    
\let\tie=\ta                    
\let\br=\ub                     


\def\a{\alpha}
\def\b{\beta}
\def\c{\chi}
\def\d{\delta}
\def\e{\epsilon}
\def\f{\phi}
\def\g{\gamma}
\def\h{\eta}
\def\i{\iota}
\def\j{\psi}
\def\k{\kappa}
\def\l{\lambda}
\def\m{\mu}
\def\n{\nu}
\def\o{\omega}
\def\p{\pi}
\def\q{\theta}
\def\r{\rho}
\def\s{\sigma}
\def\t{\tau}
\def\u{\upsilon}
\def\x{\xi}
\def\z{\zeta}
\def\D{\Delta}
\def\F{\Phi}
\def\G{\Gamma}
\def\J{\Psi}
\def\L{\Lambda}
\def\O{\Omega}
\def\P{\Pi}
\def\Q{\Theta}
\def\S{\Sigma}
\def\U{\Upsilon}
\def\X{\Xi}


\def\ve{\varepsilon}
\def\vf{\varphi}
\def\vr{\varrho}
\def\vs{\varsigma}
\def\vq{\vartheta}


\def\ca{{\cal A}}
\def\cb{{\cal B}}
\def\cc{{\cal C}}
\def\cd{{\cal D}}
\def\ce{{\cal E}}
\def\cf{{\cal F}}
\def\cg{{\cal G}}
\def\ch{{\cal H}}
\def\ci{{\cal I}}
\def\cj{{\cal J}}
\def\ck{{\cal K}}
\def\cl{{\cal L}}
\def\cm{{\cal M}}
\def\cn{{\cal N}}
\def\co{{\cal O}}
\def\cp{{\cal P}}
\def\cq{{\cal Q}}
\def\car{{\cal R}}
\def\cs{{\cal S}}
\def\ct{{\cal T}}
\def\cu{{\cal U}}
\def\cv{{\cal V}}
\def\cw{{\cal W}}
\def\cx{{\cal X}}
\def\cy{{\cal Y}}
\def\cz{{\cal Z}}


\def\Sc#1{{\hbox{\sc #1}}}      
\def\Sf#1{{\hbox{\sf #1}}}      



\def\slpa{\slash{\pa}}                            
\def\slin{\SLLash{\in}}                                   
\def\bo{{\raise-.3ex\hbox{\large$\Box$}}}               
\def\cbo{\Sc [}                                         
\def\pa{\partial}                                       
\def\de{\nabla}                                         
\def\dell{\bigtriangledown}                             
\def\su{\sum}                                           
\def\pr{\prod}                                          
\def\iff{\leftrightarrow}                               
\def\conj{{\hbox{\large *}}}                            
\def\ltap{\raisebox{-.4ex}{\rlap{$\sim$}} \raisebox{.4ex}{$<$}}   
\def\gtap{\raisebox{-.4ex}{\rlap{$\sim$}} \raisebox{.4ex}{$>$}}   
\def\TH{{\raise.2ex\hbox{$\displaystyle \bigodot$}\mskip-4.7mu \llap H \;}}
\def\face{{\raise.2ex\hbox{$\displaystyle \bigodot$}\mskip-2.2mu \llap {$\ddot
        \smile$}}}                                      
\def\dg{\sp\dagger}                                     
\def\ddg{\sp\ddagger}                                   

\font\tenex=cmex10 scaled 1200


\def\sp#1{{}^{#1}}                              
\def\sb#1{{}_{#1}}                              
\def\oldsl#1{\rlap/#1}                          
\def\slash#1{\rlap{\hbox{$\mskip 1 mu /$}}#1}      
\def\Slash#1{\rlap{\hbox{$\mskip 3 mu /$}}#1}      
\def\SLash#1{\rlap{\hbox{$\mskip 4.5 mu /$}}#1}    
\def\SLLash#1{\rlap{\hbox{$\mskip 6 mu /$}}#1}      
\def\PMMM#1{\rlap{\hbox{$\mskip 2 mu | $}}#1}   %
\def\PMM#1{\rlap{\hbox{$\mskip 4 mu ~ \mid $}}#1}       %
\def\Tilde#1{\widetilde{#1}}                    
\def\Hat#1{\widehat{#1}}                        
\def\Bar#1{\overline{#1}}                       
\def\sbar#1{\stackrel{*}{\Bar{#1}}}             
\def\bra#1{\left\langle #1\right|}              
\def\ket#1{\left| #1\right\rangle}              
\def\VEV#1{\left\langle #1\right\rangle}        
\def\abs#1{\left| #1\right|}                    
\def\leftrightarrowfill{$\mathsurround=0pt \mathord\leftarrow \mkern-6mu
        \cleaders\hbox{$\mkern-2mu \mathord- \mkern-2mu$}\hfill
        \mkern-6mu \mathord\rightarrow$}
\def\dvec#1{\vbox{\ialign{##\crcr
        \leftrightarrowfill\crcr\noalign{\kern-1pt\nointerlineskip}
        $\hfil\displaystyle{#1}\hfil$\crcr}}}           
\def\dt#1{{\buildrel {\hbox{\LARGE .}} \over {#1}}}     
\def\dtt#1{{\buildrel \bullet \over {#1}}}              
\def\der#1{{\pa \over \pa {#1}}}                
\def\fder#1{{\d \over \d {#1}}}                 


\def\frac#1#2{{\textstyle{#1\over\vphantom2\smash{\raise.20ex
        \hbox{$\scriptstyle{#2}$}}}}}                   
\def\half{\frac12}                                        
\def\sfrac#1#2{{\vphantom1\smash{\lower.5ex\hbox{\small$#1$}}\over
        \vphantom1\smash{\raise.4ex\hbox{\small$#2$}}}} 
\def\bfrac#1#2{{\vphantom1\smash{\lower.5ex\hbox{$#1$}}\over
        \vphantom1\smash{\raise.3ex\hbox{$#2$}}}}       
\def\afrac#1#2{{\vphantom1\smash{\lower.5ex\hbox{$#1$}}\over#2}}    
\def\partder#1#2{{\partial #1\over\partial #2}}   
\def\parvar#1#2{{\d #1\over \d #2}}               
\def\secder#1#2#3{{\partial^2 #1\over\partial #2 \partial #3}}  
\def\on#1#2{\mathop{\null#2}\limits^{#1}}               
\def\bvec#1{\on\leftarrow{#1}}                  
\def\oover#1{\on\circ{#1}}                              

\def\[{\lfloor{\hskip 0.35pt}\!\!\!\lceil}
\def\]{\rfloor{\hskip 0.35pt}\!\!\!\rceil}
\def\Lag{{\cal L}}
\def\du#1#2{_{#1}{}^{#2}}
\def\ud#1#2{^{#1}{}_{#2}}
\def\dud#1#2#3{_{#1}{}^{#2}{}_{#3}}
\def\udu#1#2#3{^{#1}{}_{#2}{}^{#3}}
\def\calD{{\cal D}}
\def\calM{{\cal M}}

\def\szet{{${\scriptstyle \b}$}}
\def\ulA{{\un A}}
\def\ulM{{\underline M}}
\def\cdm{{\Sc D}_{--}}
\def\cdp{{\Sc D}_{++}}
\def\vTheta{\check\Theta}
\def\fracm#1#2{\hbox{\large{${\frac{{#1}}{{#2}}}$}}}
\def\ha{{\fracmm12}}
\def\tr{{\rm tr}}
\def\Tr{{\rm Tr}}
\def\itrema{$\ddot{\scriptstyle 1}$}
\def\ula{{\underline a}} \def\ulb{{\underline b}} \def\ulc{{\underline c}}
\def\uld{{\underline d}} \def\ule{{\underline e}} \def\ulf{{\underline f}}
\def\ulg{{\underline g}}
\def\items#1{\\ \item{[#1]}}
\def\ul{\underline}
\def\un{\underline}
\def\fracmm#1#2{{{#1}\over{#2}}}
\def\footnotew#1{\footnote{\hsize=6.5in {#1}}}
\def\low#1{{\raise -3pt\hbox{${\hskip 0.75pt}\!_{#1}$}}}

\def\Dot#1{\buildrel{_{_{\hskip 0.01in}\bullet}}\over{#1}}
\def\dt#1{\Dot{#1}}

\def\DDot#1{\buildrel{_{_{\hskip 0.01in}\bullet\bullet}}\over{#1}}
\def\ddt#1{\DDot{#1}}

\def\DDDot#1{\buildrel{_{_{\hskip 0.01in}\bullet\bullet\bullet}}\over{#1}}
\def\dddt#1{\DDDot{#1}}

\def\DDDDot#1{\buildrel{_{_{\hskip 
0.01in}\bullet\bullet\bullet\bullet}}\over{#1}}
\def\ddddt#1{\DDDDot{#1}}

\def\Tilde#1{{\widetilde{#1}}\hskip 0.015in}
\def\Hat#1{\widehat{#1}}


\newskip\humongous \humongous=0pt plus 1000pt minus 1000pt
\def\caja{\mathsurround=0pt}
\def\eqalign#1{\,\vcenter{\openup2\jot \caja
        \ialign{\strut \hfil$\displaystyle{##}$&$
        \displaystyle{{}##}$\hfil\crcr#1\crcr}}\,}
\newif\ifdtup
\def\panorama{\global\dtuptrue \openup2\jot \caja
        \everycr{\noalign{\ifdtup \global\dtupfalse
        \vskip-\lineskiplimit \vskip\normallineskiplimit
        \else \penalty\interdisplaylinepenalty \fi}}}
\def\li#1{\panorama \tabskip=\humongous                         
        \halign to\displaywidth{\hfil$\displaystyle{##}$
        \tabskip=0pt&$\displaystyle{{}##}$\hfil
        \tabskip=\humongous&\llap{$##$}\tabskip=0pt
        \crcr#1\crcr}}
\def\eqalignnotwo#1{\panorama \tabskip=\humongous
        \halign to\displaywidth{\hfil$\displaystyle{##}$
        \tabskip=0pt&$\displaystyle{{}##}$
        \tabskip=0pt&$\displaystyle{{}##}$\hfil
        \tabskip=\humongous&\llap{$##$}\tabskip=0pt
        \crcr#1\crcr}}


\def\eV{\,{\rm eV}}
\def\keV{\,{\rm keV}}
\def\MeV{\,{\rm MeV}}
\def\GeV{\,{\rm GeV}}
\def\TeV{\,{\rm TeV}}
\def\sv{\left<\sigma v\right>}
\def\({\left(}
\def\){\right)}
\def\cm{{\,\rm cm}}
\def\K{{\,\rm K}}
\def\kpc{{\,\rm kpc}}
\def\beq{\begin{equation}}
\def\eeq{\end{equation}}
\def\bea{\begin{eqnarray}}
\def\eea{\end{eqnarray}}


\newcommand{\be}{\begin{equation}}
\newcommand{\ee}{\end{equation}}
\newcommand{\nbe}{\begin{equation*}}
\newcommand{\nee}{\end{equation*}}

\newcommand{\fr}{\frac}
\newcommand{\lb}{\label}

\thispagestyle{empty}

{\hbox to\hsize{
\vbox{\noindent February 2023 \hfill IPMU23-0005}
\noindent  \hfill }

\noindent
\vskip2.0cm
\begin{center}

{\large\bf Fitting power spectrum of scalar perturbations for primordial black hole production during inflation}

\vglue.3in

Daniel Frolovsky~${}^{a}$ and Sergei V. Ketov~${}^{a,b,c,\#}$ 
\vglue.3in

${}^a$~Interdisciplinary Research Laboratory, Tomsk State University\\
36 Lenin Avenue, Tomsk 634050, Russia\\
${}^b$~Department of Physics, Tokyo Metropolitan University\\
1-1 Minami-ohsawa, Hachioji-shi, Tokyo 192-0397, Japan \\
${}^c$~Kavli Institute for the Physics and Mathematics of the Universe (WPI)
\\The University of Tokyo Institutes for Advanced Study,  \\ Kashiwa 277-8583, Japan\\
\vglue.1in

${}^{\#}$~ketov@tmu.ac.jp
\end{center}

\vglue.3in

\begin{center}
{\Large\bf Abstract}  
\end{center}
\vglue.2in

\noindent We propose a simple analytic  fit for the power spectrum of scalar (curvature) perturbations during inflation, in order to describe slow roll of inflaton and formation of primordial black holes in the early universe, in the framework of single-field models. Our fit is given by a sum of the power spectrum in the slow-roll approximation, needed for a viable description of the cosmic microwave background radiation in agreement with Planck/BICEP/Keck measurements, and the log-normal (Gaussian) fit for the power spectrum enhancement (peak) needed for efficient production of primordial black holes. We use the T-type $\alpha$-attractor models in order to describe slow-roll inflation. Demanding the location and height of the peak to yield the masses of primordial black holes in the asteroid-size window allowed for the whole (current) dark matter to be composed of the primordial black holes, we find the restrictions on the remaining parameters and, most notably, on the width of the peak.

\newpage

\section{Introduction}

The inflationary paradigm was initially proposed as a possible solution to the internal problems of the standard 
(Einstein-Friedmann) cosmology such as the horizon problem, the flatness problem and the problem of initial conditions
\cite{Guth:1980zm,Linde:1981mu}. It was later recognized that inflation in the early universe may be a solution to the structure formation problem also  \cite{Liddle:2000cg}. A major recognition of the inflationary paradigm came with its success in explaining the inhomogeneity and anisotropy of the cosmic microwave background (CMB) radiation \cite{Mukhanov:2005sc}.

The underlying physics of inflation is still unknown but there is no shortage of theoretical models of inflation. The simplest single-field models of chaotic inflation are based on  the quintessence  (scalar-tensor gravity) or the modified $F(R)$-gravity theories.
More recently, the quintessence models were further generalized by adding a near-inflection point to the inflaton potential below
the inflationary scale, leading to a peak in the power spectrum of scalar perturbations that later collapse to primordial black holes (PBH) \cite{Garcia-Bellido:2017mdw,Germani:2017bcs,Germani:2018jgr,Bhaumik:2019tvl}.~\footnote{See also Refs.~\cite{Sasaki:2018dmp,Carr:2020gox,Escriva:2022duf,Karam:2022nym} and the references therein for observational constraints on PBH and their formation in single-field inflationary models.} PBH are also considered as a  good (non-particle) candidate for the present dark matter~\cite{Barrow:1992hq,Garcia-Bellido:1996mdl}. 

Usually, one begins with a particular inflationary model having a specific scalar potential, and then one numerically derives  the power spectrum by using the Mukhanov-Sasaki equation  \cite{Mukhanov:1985rz,Sasaki:1986hm}. In this paper, we begin with analytic modeling of the power spectrum of scalar (curvature) perturbations for possible PBH production in agreement with CMB measurements. We choose the simplest fit as a sum of the CMB power spectrum in the slow-roll approximation and the log-normal shape of the peak. This allows us to get analytic smooth sewing of both spectra with the minimal number of parameters and a possibility to analytically explore the whole parameter space, which is often difficult in a numerical approach. As regards the CMB power spectrum, we describe it with the help of the T-type $\a$-attractor models of inflation \cite{Kallosh:2013hoa,Galante:2014ifa} in order to get the simplest form of the spectrum during slow roll. As a result, we find new restrictions on the parameters of PBH production.
 
Our paper is organized as follows. In Sec.~2 we review the T-type $\a$-attractor models of inflation, which are used as the baseline models for CMB in the next Sections. The power spectrum of scalar perturbations during inflation in the slow-roll approximation (relevant to CMB) is derived in Sec.~3. Section 4 is devoted to our fit of the power spectrum of scalar perturbations for both CMB and PBH production, and the related spectrum of induced gravitational waves (GW).
 Our conclusion is Sec.~5.

\section{Single-field models of slow-roll inflation for CMB}

As the baseline models of large-single-field inflation, described by the standard quintessence action 
\begin{equation} \lb{quintA}
S[g_{\m\n},\f] =  \int d^4x\, \sqrt{-g}\left\{ \fracmm{M^2_{\rm Pl}}{2}R
-\fracmm{1}{2}g^{\m\n}\pa_{\m}\varphi\pa_{\n}\varphi -V(\varphi) \right\} ~~,
\end{equation} 
we choose the T-type $\a$-attractors \cite{Kallosh:2013hoa,Galante:2014ifa} with the canonical inflaton potential
\be \lb{alphaV}
V(\varphi)=V_0 \tanh^2 \left(\fracmm{\varphi/M_{\rm Pl}}{\sqrt{6\a}}\right)\equiv V_0 r^2~,\quad 
r=\tanh \fracmm{\varphi/M_{\rm Pl}}{\sqrt{6\a}}~,
\ee
where the constant $V_0$ specifies the scale of inflation, and the $\a>0$ is the free parameter of the order one. 

This model  is a viable model of large-field slow-roll inflation with a nearly flat potential, whose inflationary solution is an attractor describing chaotic inflation, being very close to the Starobinsky model \cite{Starobinsky:1980te} in the case of $\a=1$.
The CMB tilt of scalar perturbations, predicted by the T-model is given by the simple formula \cite{Mukhanov:1981xt} 
\be \lb{mc}
n_s = 1 - \fracmm{2}{N_e}~,
\ee
in terms of e-folds $N_e$ as the running variable describing time evolution, $N_e(k)=\ln(k_{\rm final}/k)$ as the function of scale $k$ \cite{Hodges:1990bf}. The CMB tensor-to-scalar ratio $r$ is approximately $(N_e\gg 1)$ given by 
\cite{Kallosh:2013hoa,Galante:2014ifa} 
\be \lb{ralpha}
r_{\rm \a} \approx \fracmm{12\a}{N_e^2}~,
\ee
providing the comfortable theoretical prediction against future measurements of $r$.  Indeed, the current CMB measurements by Planck/BICEP/Keck collaborations \cite{Planck:2018jri,BICEP:2021xfz,Tristram:2021tvh} give
\be \lb{ns}
n_s = 0.9649 \pm 0.0042 \quad (68\%~{\rm C.L.})~,\quad r < 0.036\quad (95\%~{\rm C.L.})~~,
\ee
while they are in good agreement with Eqs.~(\ref{mc}) and (\ref{ralpha}) with the best fit close to $N_{e}=55$.

The T-model potential (\ref{alphaV}) is symmetric under the sign change $\varphi\to -\varphi$ but is not periodic. Periodicity
of inflaton potential is assumed in the models of natural inflation and PBH production, where inflaton is identified with an axion, see e.g., Ref.~\cite{Gao:2020tsa}. The periodicity in the $\a$-attractor models can be achieved via changing the global shape of the potential without affecting slow roll inflation by replacing the function $\tanh\f$ in Eq.~(\ref{alphaV}) by 
 the periodic (Jacobi) elliptic function  $sn (\f\left|\right.k)$ with the elliptic modulus $0<k^2< 1$, due to the known approximation 
\be \lb{jacobi} 
sn(\f \left|\right. k) \approx \tanh \f \quad {\rm for} \quad k^2\to 1~~, \quad \f=\fracmm{\varphi/M_{\rm Pl}}{\sqrt{6\a}}~,
\ee  
thus combining the theoretically attractive features of  chaotic inflation and natural inflation.

The generalization of the simplest T-model potential (\ref{alphaV}) to the form \cite{Kallosh:2013hoa,Galante:2014ifa} 
\be \lb{tpot}
 V_{\rm gen.}(\varphi)=f^2\left(\tanh \fracmm{\varphi/M_{\rm Pl}}{\sqrt{6\a}}\right)
 \ee
with a monotonically increasing (during slow roll) function $f(r)$, $r=\tanh \fracmm{\varphi/M_{\rm Pl}}{\sqrt{6\a}}$, can be used
for engineering a near-inflection point in the potential, leading to a peak (enhancement) in the power spectrum of scalar
perturbations, needed for PBH formation \cite{Iacconi:2021ltm,Braglia:2022phb}.~\footnote{The generalizations
of the Starobinsky model and the E-type $\a$-attractors, accommodating a near-inflection point for  PBH production, were proposed in Refs.~\cite{Frolovsky:2022ewg} and \cite{Frolovsky:2022qpg}, respectively.}

In the generalized T-models (\ref{tpot}) slow-roll inflation occurs for large positive values of the inflation field $\varphi$ with an approximate scalar potential  of the E-type \cite{Dalianis:2018frf} as  $(M_{\rm Pl}=1)$
\be \lb{asymptal} 
V(\varphi) = f_{\infty}^2  - 4f_{\infty}{f'}_{\infty} \, e^{-\sqrt{\fracmm{2}{3\a}}\varphi} +{\cal O}
\left( e^{-2\sqrt{\fracmm{2}{3\a}}\varphi} \right)~,
\ee
where we have introduced the parameters $f_{\infty}=\left. f\right|_{\varphi\to\infty}$ and
${f'}_{\infty}=\left. \pa_{\varphi}f\right|_{\varphi\to\infty}$. The constant in front of the second term in Eq.~(\ref{asymptal}) 
can be chosen at will by a constant shift of the inflaton field $\varphi$, so that the potential (\ref{asymptal}) can be simplified to 
\be \lb{as2} 
V(\varphi) = V_0\left( 1  - e^{-\sqrt{\fracmm{2}{3\a}}\varphi}\right) +{\cal O}\left( e^{-2\sqrt{\fracmm{2}{3\a}}\varphi} \right)~,
\ee
which implies Eqs.~(\ref{mc}) and (\ref{ralpha}). 

The $\a$-attractors with $\a\neq 1$ do not have a simple description on the dual $F(R)$-gravity side, see e.g.,
Ref.~\cite{Ivanov:2021chn} for details of the correspondence. The Starobinsky function $F(R)=\fracmm{M^2_{\rm Pl}}{2}(R+\fracmm{R^2}{6m_{\rm inf.}^2})$ on the modified gravity side arises in the case of $\a=1$ and $f(r) =\sqrt{3}m_{\rm inf.}M_{\rm Pl}r/(r+1)$, where $m_{\rm inf.}$ is the inflaton (scalaron) mass. In general, the exact dual $F(R)$ gravity function associated with any inflaton potential $V$ in the model (\ref{quintA}) is only known in the parametric (implicit) form, see Eqs.~(2.7) and (2.8) in Ref.~\cite{Ivanov:2021chn}, as
\begin{eqnarray} 
R &=& \left[\fracmm{\sqrt{6}}{M_{\rm Pl}}V_{,\varphi}+\fracmm{4V}{M^2_{Pl}}\right]
\exp\left( \sqrt{\fracmm{2}{3}}\fracmm{\varphi}{M_{Pl}}\right)~~,
\label{RV}\\
F &=& \fracmm{M^2_{\rm Pl}}{2}\left[\fracmm{\sqrt{6}}{M_{\rm Pl}}V_{,\varphi}+\fracmm{2V}{M^2_{\rm Pl}}
\right] \exp\left( 2\sqrt{\fracmm{2}{3}}\fracmm{\varphi}{M_{\rm Pl}}\right)~.
\label{FV}
\end{eqnarray}
When $\a\neq 1$, or $A(R)\neq const.$ in the slow-roll approximation with the potential (\ref{alphaV}), we find that the $F$-function can be approximated in the form
\be \lb{slowvary} F(R) = \fracmm{M^2_{\rm Pl}}{2}\left[ R+A(R)\fracmm{R^2}{6m^2}\right]
\ee
with the function 
\be \lb{slowAT}
A(R) \approx 1 - \fracmm{3}{4}\ve~,
\ee
where $\ve$ is the standard slow-roll parameter 
\be \lb{esroll}
 \ve = \fracmm{M^2_{\rm Pl}}{2} \left(\fracmm{V_{,\varphi}}{V}\right)^2~~,
\ee
and $R\approx 12H^2\approx 4V/M^2_{\rm Pl}$ in terms of the Hubble function $H$ and the potential $V$.
 
The particular examples of the generalized T-models, suitable for inflation and PBH production, can be obtained by expanding
the $f(r)$-function in Taylor series and tuning the expansion coefficients \cite{Dalianis:2018frf,Iacconi:2021ltm}.

The slow-roll evolution of inflaton with e-folds $N$ as the running (time) variable is described by the (non-linear) equation of motion,
obtained from the standard (Klein-Gordon) equation minimally coupled to gravity in the (spatially flat) universe, when the acceleration term is ignored,
\be \lb{chva}
\fracmm{1}{M^2_{\rm Pl}} \left(\fracmm{d\varphi}{dN}\right)^2=\fracmm{d \ln V}{dN}~.
\ee 
This equation has an {\it exact} solution in the case of the T-model potential (\ref{alphaV}), with
\be \lb{esol}
\varphi/M_{\rm Pl}=\sqrt{2N_0} ~{\rm arcosh} \left(\fracmm{N}{N_0}\right)~, \quad N \gg N_0 >0~,
\ee
where the (implicit) integration constant is associated with constant shifts of the field $\varphi$. 
The solution implies
\be \lb{Tsol}
\fracmm{N-N_0}{N+N_0} = \tanh^2\left(\fracmm{\varphi/M_{\rm Pl}}{\sqrt{6\a}}\right)~,\quad 
N_0 = \fracmm{3\a}{4}~~,
\ee
and gives a very simple potential $V(N)$ of the T-model in the slow-roll approximation,
\be \lb{TpotN}
V(N) = V_0 \left( \fracmm{N-N_0}{N+N_0} \right)~~.
\ee

The Hubble function $H(N)$ is also simply related to the potential $V(H)$ is the slow-roll approximation via the Friedmann equation
\be \lb{Fried}
H^2(N) = \fracmm{V(N)}{3M^2_{\rm Pl}}~.
\ee

The relations between the potential $V(N)$, the running tensor-to-scalar ratio $r(N)$ and the slow-roll parameter $\ve(N)$  in the slow-roll approximation are very simple too,
\be \lb{rnew}
 r(N) = 16\ve (N) = 8 \fracmm{d\ln V}{dN}=\fracmm{12\a}{N^2-N_0^2}= \fracmm{12\a}{N^2-(3\a/4)^2}~,
\ee
leading to a bit more precise formula than Eq.~({\ref{ralpha}).

The very simple form (\ref{TpotN}) of the T-potential $V(N)$ in the slow-roll approximation is one of the reasons why we choose the T-models as our baseline models in this paper.

\section{Power spectrum of scalar perturbations in slow-roll approximation}

Primordial scalar perturbations ($\z$) and primordial  tensor perturbations (primordial gravitational waves
$g$) are defined by a perturbed Friedmann-Lemaitre-Robertson-Walker (FLRW) metric, 
\be \lb{perm}
ds^2 = dt^2 - a^2(t)\left( \d_{ij}+ h_{ij}(\vec{r})\right) dx^idx^j~~,\qquad i,j=1,2,3~,
\ee
where
\be \lb{pertur}
h_{ij}(\vec{r}) =2\z(\vec{r})\d_{ij} + \sum_{a=1,2} g^{(a)}(\vec{r})e^{(a)}_{ij}(\vec{r})~~,\quad H=\fracmm{da/ dt}{a}~,
\ee
in terms of the local basis $e^{(a)}$ obeying the relations $e^{i(a)}_{i}=0$, 
 $g^{(a)}_{,j}e^{j(a)}_{i}=0$ and $e^{(a)}_{ij}e^{ij(a)}=1$.

The primordial spectrum $P_{\z}(k)$ of scalar (density) perturbations  is defined by the 2-point correlator of scalar perturbations,
\be \lb{powersp}
\VEV{\z^2(\vec{r})}=\int dk \fracmm{P_{\z}(k)}{k}~.
\ee
The CMB power spectrum can be described by the Harrison-Zeldovich fit
\be \lb{hz} P^{\rm HZ}_{\z}(k)\approx 2.21^{+0.07}_{-0.08} \times 10^{-9}\left(\fracmm{k}{k_*}\right)^{n_s-1}~~,
\ee
near the pivot scale $k_*=0.05~{\rm Mpc}^{-1}$, or in the slow-roll (SR) approximation by 
\be \lb{slP}
P^{\rm SR}_{\z}(k)\approx P_0\ln^2{\left(\fracmm{k}{k_{\text{final}}}\right)}~, \quad P_0=const.
\ee

The power spectrum $P_{\z}(N)$ is simply related to the potential $V(N)$ in the slow-roll approximation via the standard relation, see e.g., Refs.~\cite{Garcia-Bellido:2017mdw,Frolovsky:2022qpg},
\be \lb{powerV}
P_{\z}(N) = \fracmm{V^2}{12\pi^2M^4_{\rm Pl}}\left( \fracmm{dV}{dN}\right)^{-1}~~.
\ee
It also implies 
\be \lb{nsP}
1-n_s =  \fracmm{d\ln P_{\z}(N)}{dN}~.
\ee

In the case of the potential (\ref{TpotN}), we find  very simple equations,
\be  \lb{psT}
P^{\rm SR}_{\z}(N) = \fracmm{V_0}{18\pi^2M^4_{\rm Pl}\a} (N-N_0)^2\equiv P_0 (N-N_0)^2~,
\ee
and
\be \lb{nsT}
n_s = 1- \fracmm{2}{N-N_0}~~,
\ee
where the last equation reproduces Eq.~(\ref{mc}).

The observed CMB window into inflation does not allow us to reconstruct the full inflaton scalar potential from the power spectrum beyond the slow-roll region. The well-known reconstruction formula, proposed by Hodges and Blumentahl \cite{Hodges:1990bf} in the form
\be \lb{hb1}
\fracmm{1}{V(N)} = - \fracmm{1}{12\pi^2M^4_{\rm Pl}}\int \fracmm{dN}{P_{\z}(N)}~,
\ee
requires knowing the full power spectrum at different scales and the limits of integration. Moreover, the reconstruction procedure
should be based on getting exact solutions to the Mukhanov-Sasaki equation instead of the slow-roll solution in Eq.~(\ref{psT}), see e.g., Ref.~\cite{Dudas:2012vv} for some examples. It is not our purpose in this paper to reconstruct the inflaton potential beyond its qualitative features. Nevertheless, it may be possible for the CMB region under some additional assumptions, e.g., when assuming a very low value of the tensor-to-scalar ratio $r$, which  implies a small correction $\d V$ to the constant $V_0$ defining the inflationary scale, i.e. the inflaton potential in the form $V=V_0+\d V$ with  $\abs{\d V}\ll V_0$. Then Eq.~(\ref{hb1}) is simplified to
\be \lb{hb2}
\d V(N) = \fracmm{V_0}{12\pi^2M^4_{\rm Pl}}\int \fracmm{dN}{P_{\z}(N)}~,
\ee
while the integration constants merely rescale $V_0$ and shift $N$. Equation (\ref{chva}) also gets simplified to
\be \lb{chvaS}
\fracmm{V_0}{M^2_{\rm Pl}} \left(\fracmm{d\varphi}{dN}\right)^2=\fracmm{d (\d V)}{dN}~~.
\ee 
Then a partial reconstruction of the scalar potential becomes possible from the CMB power spectrum $P_{\z}(N)$ of scalar perturbations without knowing the power spectrum of tensor perturbations, which is also true for the $\a$-attractors when the parameter $\a$ is small enough, $\a\leq 1$, with
\be  \lb{reconV}
V(\varphi) \approx V_0 \left( 1  - e^{-\sqrt{\fracmm{2}{3\a}}\varphi}\right)~~,
\ee
as in Eq.~(\ref{as2}). This is yet another reason for us to take the T-models of $\a$-attractors as our baseline models of inflation and generalize their power spectrum of scalar perturbations by adding a peak at higher values of $k$. 

\section{Log-normal fit for a peak and GW spectrum}

The log-normal fit is the simplest (Gaussian) description of a peak in the power spectrum, see  e.g., Ref.~\cite{Pi:2020otn}. A power-law ansatz for the peak was considered in Ref.~\cite{Hertzberg:2017dkh}. In this paper we propose another ansatz for the power spectrum, combining the CMB spectrum in the slow-roll approximation with the log-normal fit for the enhancement (peak) of the power spectrum needed for PBH formation at a lower scale,
\begin{equation} \lb{psfit}
    \displaystyle P_{\z}(k)= P_0\ln^2{\left(\fracmm{k}{k_{\text{final}}}\right)}+ A\fracmm{\exp{\displaystyle\Big[-\fracmm{\ln^2{\left(\frac{k}{k_{\rm peak}}\right)}}{2 \sigma ^2}}\Big]}{\sqrt{2 \pi } \sigma }~~,
\end{equation}
where $k_{\rm peak}$ is a position of the peak,  $\sigma>0$  is the width of the peak  and $A$ is the normalization of the peak amplitude, $A\approx (\sqrt{2\pi}\sigma) 0.01$, needed for efficient PBH production (about $10^7$ higher than the CMB amplitude).
The normalization factor $P_0$ is given by Eq.~(\ref{psT}), 
\begin{equation}
    P_0=\fracmm{V_0}{18\pi^2 \alpha M_{\text{Pl}}^4}~~,
\end{equation}
see also Eq.~(\ref{slP}). When $V_0 \sim m_{\rm inf.}^2 M_{\text{Pl}}^2$, $ m_{\rm inf.}\sim 10^{-5} M_{\text{Pl}}$ and  $\alpha \sim 1$, we get
$P_0\sim  \mathcal{O}(10^{-12})$~. 
The power spectrum (\ref{psfit}) can be rewritten, using the e-foldings variable $N$  via the relation 
\begin{equation} \lb{kNrel}
   d \ln{k}=-dN~,
\end{equation}
to the simple form
\begin{equation} \lb{psN}
    \displaystyle P_{\z}(N)= P_0\left(N-N_0\right)^2
    + A\fracmm{\exp{\displaystyle\Big[-\fracmm{\left(N-N_{\rm peak}\right)^2}{2 \sigma ^2}}\Big]}{\sqrt{2 \pi } \sigma }~~.
\end{equation}
We choose $k_{\text{\rm final}}=e^{N_e}$ $\text{Mpc}^{-1}$ $\approx7.7\cdot 10^{23}~\text{Mpc}^{-1}$. 

The PBH masses can be estimated by the relation \cite{Sasaki:2018dmp,Carr:2020gox,Escriva:2022duf,Karam:2022nym}
\begin{equation}
	\fracmm{M_{\rm PBH}(k)}{M_{\odot}} \simeq 10^{-16}\left(\fracmm{k}{10^{14} \mathrm{Mpc}^{-1}}\right)^{-2}~.
\end{equation}
We choose $k_{\rm peak}$ or $N_{\rm peak}$ to get $M_{\rm PBH}$ within the current observational window for PBH as the whole dark matter  \cite{Sasaki:2018dmp,Carr:2020gox,Escriva:2022duf}, i.e. between $10^{17}$ g and $10^{21}$ g. For example,
when $k_{\rm peak}\approx 10^{13}~\text{Mpc}^{-1}$, we get $M_{\rm PBH}\approx 2\cdot 10^{19}$ g. The profile of the power spectrum in given on Fig.~1 for some values of $\sigma$.

\begin{figure}[h]
\center{\includegraphics[width=0.7\textwidth]{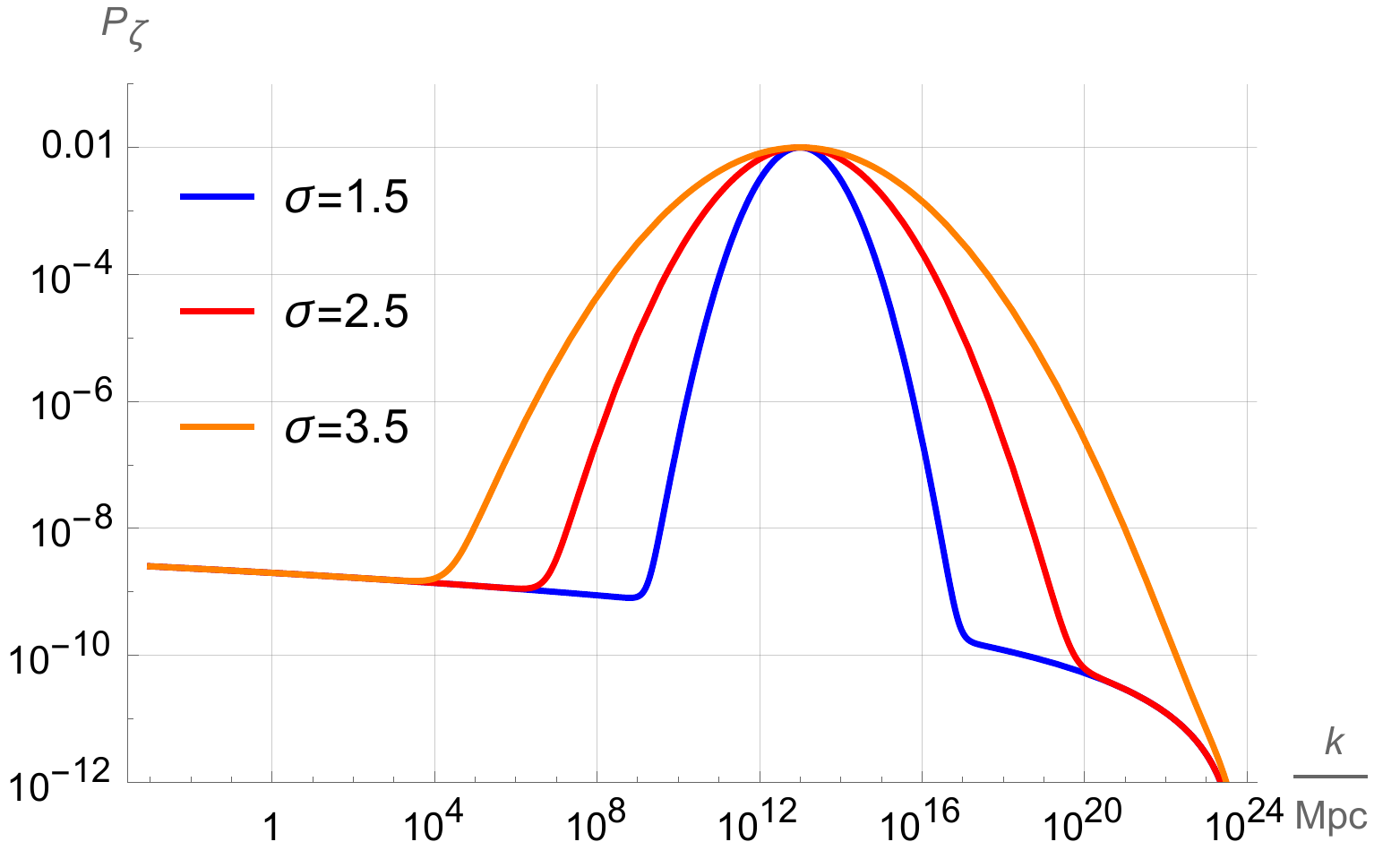}\\}
\caption{The power spectrum with the parameters $P_0=6.57\cdot 10^{-13}$, $k_{\rm peak}=10^{13}~\text{Mpc}^{-1}$, $k_{\text{\rm final}}=7.7\cdot 10^{23} $ $\text{Mpc}^{-1}$ for $\s>1$.  }
\end{figure}

Equations (\ref{nsP}) and (\ref{psN}) imply the spectral tilt
\begin{equation}\label{exns}
    n_s=1-\fracmm{2 \left(N-N_0\right)-\fracmm{A \left(N-N_{\rm peak}\right) e^{-\fracmm{\left(N-N_{\rm peak}\right){}^2}{2 \sigma ^2}}}{P_0\sqrt{2 \pi } \sigma}}{\fracmm{A e^{-\fracmm{\left(N-N_{\rm peak}\right){}^2}{2 \sigma ^2}}}{P_0\sqrt{2 \pi } \sigma }+\left(N-N_0\right){}^2}~~.
\end{equation}

It follows from Eq.~(\ref{exns}) versus Eq.~(\ref{mc}) that the tilt $n_s$ gets the exponentially small corrections (back reaction)  from the peak. 
To quantitatively evaluate an impact of the back reaction, we introduce the dimensionless parameters for the relative scales,
\be \lb{relscales}
\m_L = \fracmm{k_{\rm left}- k_*}{k_{\rm left}} \quad {\rm and} \quad \m_R = \fracmm{k_{\rm final}- k_{\rm right}}{k_{\rm final}}~~,
\ee
characterizing the separation between the CMB pivot scale $k_*$ and the left end $k_{\rm left}$ of the peak, and the separation between
the end of inflation $k_{\rm final}$ and the right end $k_{\rm right}$ of the peak, respectively. Since the CMB pivot scale and the PBH scales have to be separated,  it implies  $k_*<k_{\rm left}$ or $\m_L>0$. The exponential corrections in Eq.~(\ref{exns}) are negligible when 
$k_{\rm left}\geq 10^3~{\rm Mpc}^{-1}$. On the other hand, the right end of the peak must be within inflation, so that $\m_R>0$. We expect
$k_{\rm right}$ to be close to the end of inflation.~\footnote{Particle production is also more efficient toward the end of inflation \cite{Addazi:2017ulg}.}

We illustrate those considerations by our numerical calculations with the results displayed on Figs.~2 and 3 for various values of $k_{\rm peak}$ and $\s$ against the observed values of the tilt $n_s$ in Eq.~(\ref{ns}). The black curves in the ($k_{\rm peak},\s)$-plane correspond to the condition $\m_R=0$. The area above the black curve and the white area are forbidden.

\begin{figure}[h]
\begin{minipage}[h]{0.49\linewidth}
\center{\includegraphics[width=1\textwidth]{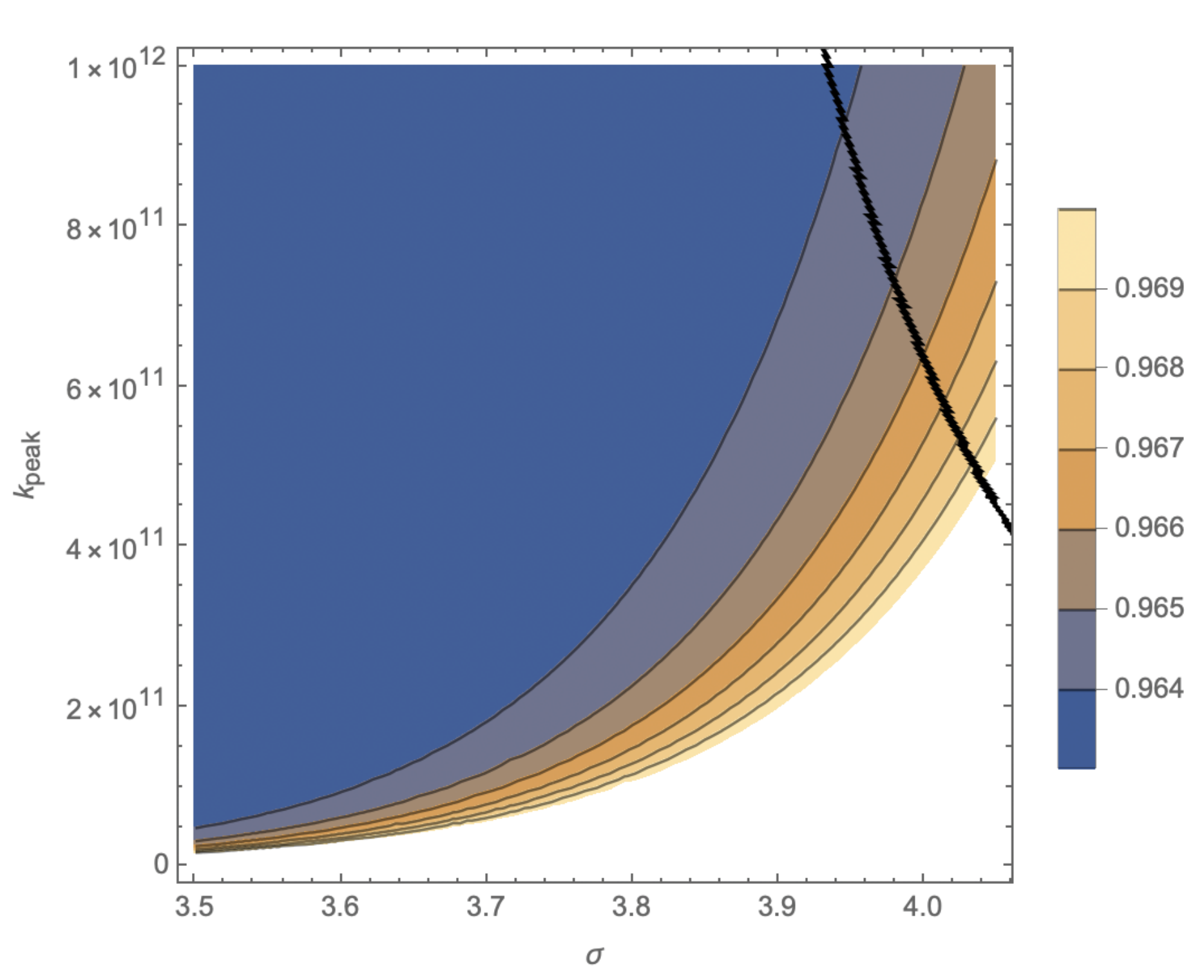}}
\end{minipage}
\hfill
\begin{minipage}[h]{0.49\linewidth}
\center{\includegraphics[width=1\textwidth]{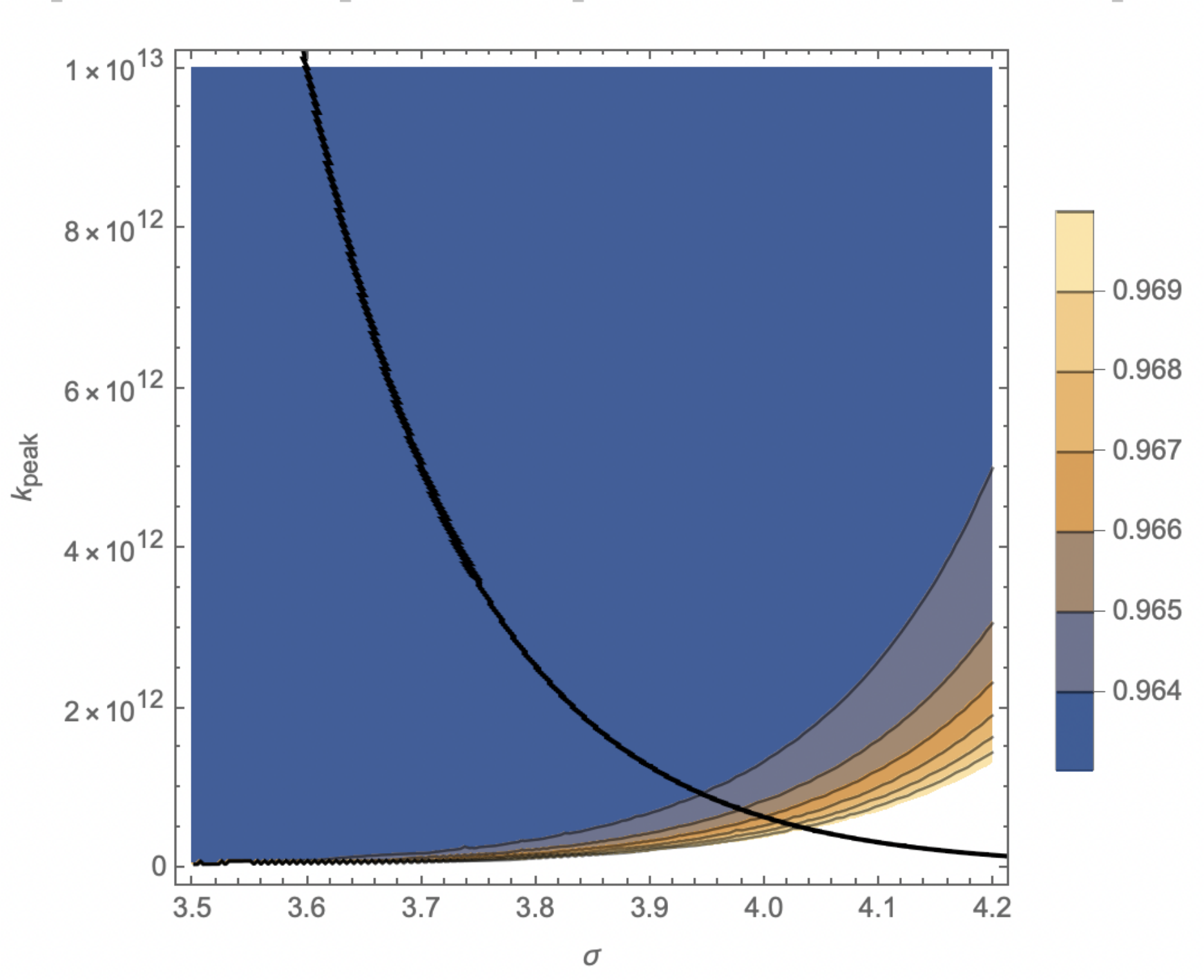}}
\end{minipage}
\caption{The impact of Eq.~(\ref{exns}) on the parameters of our model  for $10^{11}~{\rm Mpc}^{-1}k_{\rm peak}\leq 10^{12}~{\rm Mpc}^{-1}$ (on the left) and for $10^{12}~{\rm Mpc}^{-1}\leq k_{\rm peak}\leq 10^{13}~{\rm Mpc}^{-1}$ (on the right)  in the 
$(\sigma,k_{\rm peak})$-plane. The (excluded) area above the black curve leads to the right end of the peak after the end of inflation. 
The other parameters are  $P_0=6.57\cdot 10^{-13}$ and $k_{\text{\rm final}}=7.7\cdot 10^{23}~\text{Mpc}^{-1}$. }
\end{figure}
 
 \begin{figure}
\center{\includegraphics[width=0.45\textwidth]{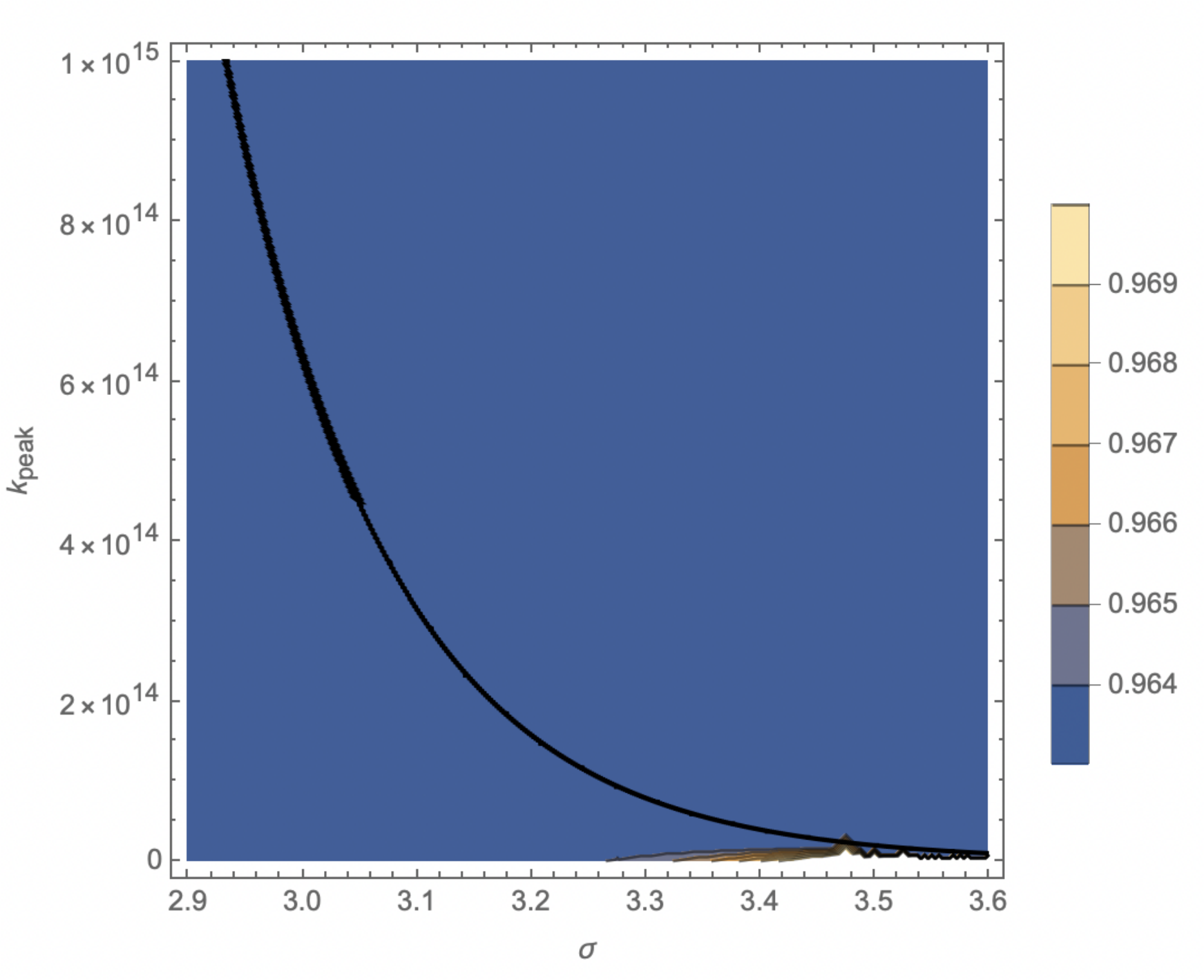}\\}
\caption{The impact of Eq.~(\ref{exns}) on the parameters for $10^{14}~{\rm Mpc}^{-1}\leq k_{\rm peak}\leq 
10^{15}~{\rm Mpc}^{-1}$ in the $(\sigma,k_{\rm peak})$-plane. The (excluded) area above the black curve leads to the right  end of the peak after the end of inflation. The other parameters are  $P_0=6.57\cdot 10^{-13}$ and $k_{\text{\rm final}}=7.7\cdot 10^{23}~\text{Mpc}^{-1}$. }
\end{figure}

For example, when $N_e=55$ and $N_0=3/4$ (or $\a=1$), we get $n_s\approx 0.9631$ from Eq.~(\ref{nsT}), whereas we get $n_s\approx 0.9649$ after taking into account the exponential terms in Eq.~(\ref{exns}) with the parameters $k_{\rm peak}=6\cdot 10^{11}~{\rm Mpc}^{-1}$ and  $\sigma=3.945$.

 Our analysis allows us to restrict (from above) the possible peak width values  $\sigma$ at fixed $k_{\rm peak}$ and duration of inflation $N_e$ (or $n_s$). We summarize those restrictions in Table 1.~\footnote{When $k_{\rm peak}>10^{15}~\text{Mpc}^{-1}$, the PBH masses are lower than the Hawking evaporation limit of $10^{15}g$ for black holes.} 
\begin{table}[h]
\begin{center}
\begin{tabular}{| c | c | c |}
\hline
\phantom{@}$M_{\rm PBH},\text{g}$\phantom{@} & \phantom{@}$k_{\rm peak},\text{Mpc}^{-1}$ \phantom{@} & $\sigma$ \\ \hline
$10^{21}$ & $1.41\cdot 10^{12}$ & $\leq 3.89$  \\ \hline
$10^{20}$ & $4.46\cdot 10^{12}$ &  $\leq 3.73$ \\ \hline
$10^{19}$ & $1.41\cdot 10^{13}$ &  $\leq 3.56$  \\ \hline
$10^{18}$ & $4.46\cdot 10^{13}$  & $\leq 3.40$  \\ \hline
$10^{17}$ & $1.41\cdot 10^{14}$  & $\leq 3.23$   \\ \hline
\end{tabular}
\caption{The PBH masses $M_{\rm PBH}$, the scales $k_{\rm peak}$ and the upper bounds on $\s$.} 
\end{center}
\end{table}

The spectrum of the induced GW can be derived by using the standard formula obtained in the second order with respect to
perturbations \cite{Domenech:2021ztg},
\begin{equation} \lb{GWsp}
\begin{array}{c}
\displaystyle \Omega_{GW}(k)=\fracmm{\Omega_{r,0}}{32}\int\limits_0^\infty dv\int\limits_{|1-v|}^{1+v}du\fracmm{\mathcal{T}(u,v)}{u^2v^2}P_{\z}(vk)P_{\z}(uk),\\[6mm]
\displaystyle \mathcal{T}(u,v)=\frac{1}{4}\Big[\fracmm{4v^2-(1+v^2-u^2)^2}{4uv}\Big]^2
\Big(\fracmm{u^2+v^2-3}{2uv}\Big)^4
\Big[\Big(\ln\Big|\fracmm{3-(u+v)^2}{3-(u-v)^2}\Big|\\[3mm]
\displaystyle -\,\fracmm{4uv}{u^2+v^2-3}\Big)^2+\pi^2\Theta\big(u+v-\sqrt{3}\big)\Big]~~,
\end{array}
\end{equation}
where $\Omega_{r,0}=8.6\cdot 10^{-5}$. Our numerical results for a wide peak with $\s>1$ are displayed on Fig.~4.

\begin{figure}[h]
\center{\includegraphics[width=0.7\textwidth]{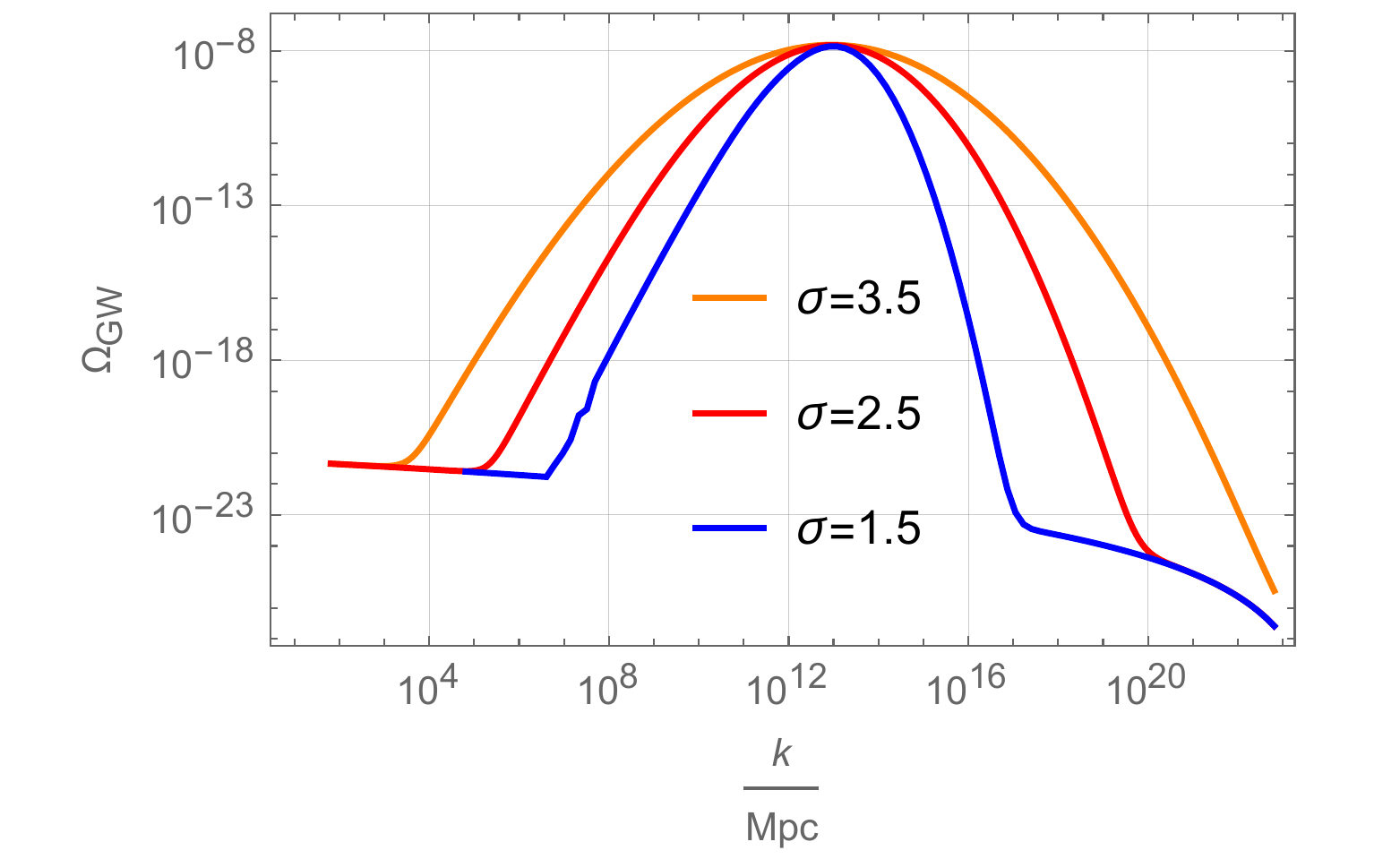}\\}
\caption{The induced GW spectrum for selected values $\s>1$ of a wide peak in the power spectrum, with the parameters  $P_0=6.57\cdot 10^{-13}$, $k_{\rm peak}=10^{13}~\text{Mpc}^{-1}$ and $k_{\text{\rm final}}=7.7\cdot 10^{23}~\text{Mpc}^{-1}$.} 
\end{figure}

The peak in the GW-spectrum associated with a wide ($\s>1$) peak in the power spectrum can be analytically approximated as 
\begin{equation} \lb{wideO}
\Omega^{{\rm (peak)}}_{{\rm GW},r} \approx 0.125 \fracmm{A^2}{\sigma^2}\exp{\left[\fracmm{-\ln^2\left(\fracmm{k}{k_{\rm peak}}\right)}{\sigma^2}\right]}\sim 10^{-6} P^2_{\z}(k)~~.
\end{equation}

For comparison, in Fig.~5 we give our numerical results for the induced GW spectrum with selected values $\s<1$ of a sharp peak in the power  spectrum. Then the GW spectrum is not given by a sum of contributions from the peak and the slow-roll, while the simple relation to the power spectrum in Eq.~(\ref{wideO}) is also not valid. Instead, the cross  terms in Eq.~(\ref{GWsp}) become significant and the shape of the GW spectrum changes, see Fig.~5. This phenomenon was also observed  in 
Ref.~\cite{Balaji:2022dbi}.

\begin{figure}[h]
\center{\includegraphics[width=0.7\textwidth]{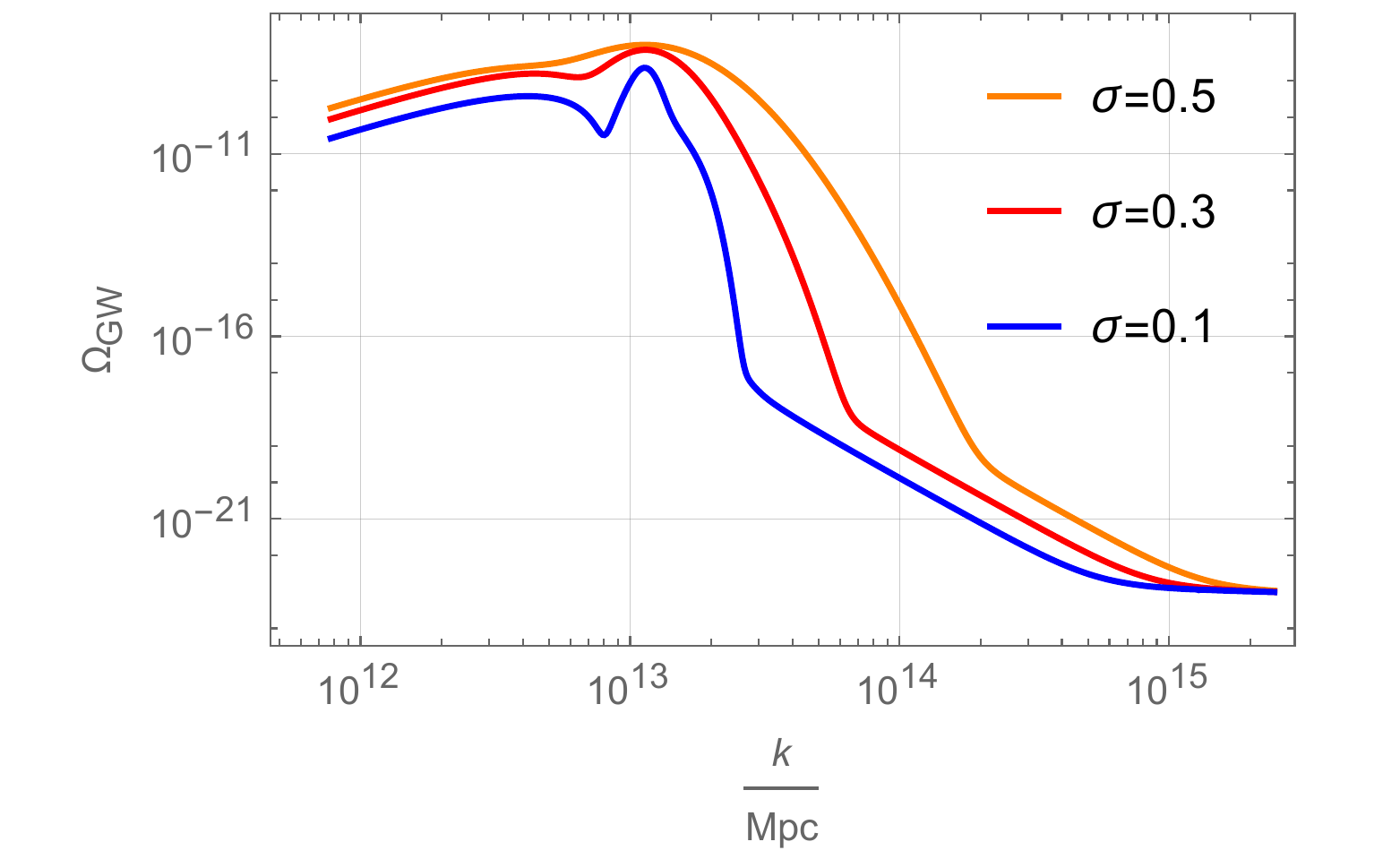}\\}
\caption{The induced GW spectrum for selected values $\s<1$ of a sharp peak in the power spectrum, with the parameters  $P_0=6.57\cdot 10^{-13}$, $k_{\rm peak}=10^{13}~\text{Mpc}^{-1}$ and $k_{\text{\rm final}}=7.7\cdot 10^{23}~\text{Mpc}^{-1}$.} 
\end{figure}

\section{Conclusion}

Our investigation in this paper is based on the ansatz (\ref{psfit}) for the power spectrum of scalar (curvature) perturbations during inflation. The ansatz is given by a sum of the CMB power spectrum in the slow-roll approximation and the log-normal fit for the power spectrum enhancement (peak) needed for efficient PBH production. The ansatz  (\ref{psfit}) is very simple, while we use the slow-roll approximation and the T-type
$\a$-attractor models of inflation in order to justify the first term in Eq.~(\ref{psfit}). The second term in Eq.~(\ref{psfit}) requires the scalar potential in those models to be generalized, e.g., via engineering a near-inflection point and an ultra-slow-roll phase during inflation, see e.g.,
Refs.~\cite{Garcia-Bellido:2017mdw,Germani:2017bcs,Karam:2022nym,Iacconi:2021ltm,Braglia:2022phb,Frolovsky:2022ewg,Frolovsky:2022qpg,Dalianis:2018frf,Ragavendra:2020sop,Balaji:2022rsy} for explicit examples. We are aware that the slow-roll approximation is  violated during the ultra-slow-roll phase needed for a peak generation, and do not expect that Eq.~(\ref{psfit}) is suitable for a full reconstruction of the inflaton scalar potential. Instead, we take the power spectrum (\ref{psfit}) for granted and study  its consequences, both analytically and numerically, in the context of CMB and PBH as DM. 

Our main results are given in Sec.~4 including our Figures and Table 1 that summarizes the restrictions on the peak  width $\s$ from above.

\section*{Acknowledgements}

This work was partially supported by Tomsk State University under the development program Priority-2030.  SVK was also supported by Tokyo Metropolitan University, the Japanese Society for Promotion of Science under the grant No.~22K03624, and the World Premier International Research Center Initiative, MEXT, Japan. 

One of the authors (SVK) is grateful to Shyam Balaji, Guillem Domenech, Noriaki Kitazawa, Laura Iacconi, Misao Sasaki and Alexei Starobinsky for discussions and correspondence.

\bibliography{Bibliography}{}
\bibliographystyle{utphys}

\end{document}